\def\bar{\begin{eqnarray}}
\def\ear{\end{eqnarray}}
\def\bb{\bibitem}
\def\eqi{\begin{linenomath*}\begin{equation}}
\def\eqf{\end{equation}\end{linenomath*}}
\def\eqia{\begin{eqnarray}}
\def\eqfa{\end{eqnarray}}
\def\rp#1#2{{#1\over#2}}
\def\ct#1{\cite{#1}}

\def\oc2{$\mathcal{O}(c^{-2})$}

\documentclass{ws-procs975x65}

\begin{document}

\wstoc{Testing DGP modified gravity in the Solar System}{L. Iorio}

\title{TESTING DGP MODIFIED GRAVITY IN THE SOLAR SYSTEM}

\author{LORENZO IORIO\footnote{Fellow of the Royal Astronomical Society}}

\address{
Viale Unit$\grave{ \it a}$ di Italia 68, 70125, Bari (BA),
Italy\\
\email{lorenzo.iorio@libero.it}}

\begin{abstract}
In this talk we review the perspectives of testing the
multidimensional Dvali-Gabadadze-Porrati (DGP) model of modified
gravity in the Solar System. The inner planets, contrary to the
giant gaseous ones, yield the most promising scenario for the near
future.
\end{abstract}

\bodymatter

\section{The DGP picture}
In the Dvali-Gabadadze-Porrati (DGP) braneworld scenario
\cite{DGP00} our Universe is a (3+1) space-time brane embedded in
a five-dimensional Minkowskian bulk. All the particles and fields
of our experience are constrained to remain on the brane apart
from gravity which is free to explore the empty bulk. Beyond a
certain threshold $r_0$, which is a free-parameter of the theory
and is fixed by observations to $\sim 5$ Gigaparsec, gravity
experiences strong modifications with respect to the usual
four-dimensional Newton-Einstein picture: they allow to explain
the observed acceleration of the expansion of the Universe without
resorting to the concept of dark energy. For a recent review of
the phenomenology of DGP cosmologies see Ref. 2. With more
details, an intermediate regime is set by the Vainshtein scale
$r_{\star} = (r_g r^2_0)^{1/3}$, where $r_g = 2GM/c^2$ is the
Schwarzschild radius of a central object of mass $M$ acting as
source of gravitational field; $G$ and $c$ are the Newtonian
gravitational constant and the speed of light in vacuum,
respectively. For a Sun-like star $r_{\star}$ amounts to about 100
parsec. In the process of recovering the 4-dimensional
Newton-Einstein gravity for $r << r_{\star} << r_0$, DGP predicts
small deviations from it which yield to effects observable at
local scales\ct{DGZ03}. They come from an extra radial
acceleration of the form\ct{Gru05, LS03, Ior05a}
\begin{equation} {\bf a}_{\rm DGP} =
\mp\left(\rp{c}{2r_0}\right)\sqrt{\rp{GM}{r}}\hat{r}.
\end{equation} The minus sign is related to a cosmological phase in
which, in absence of cosmological constant on the brane, the
Universe decelerates at late times, the Hubble parameter $H$
tending to zero as the matter dissolves on the brane: it is called
Friedmann-Lema\^{\i}tre-Robertson-Walker (FLRW) branch. The plus
sign is related to a cosmological phase in which the Universe
undergoes a de Sitter-like expansion with the Hubble parameter $H
= c/r_0$ even in absence of matter. This is the self-accelerated
branch, where the accelerated expansion of the Universe is
realized without introducing a cosmological constant on the brane.
Thus, there is a very important connection between local and
cosmological features of gravity in the DGP model.
\section{The testable effects and their measurability}
About the local effects, Lue and Starkman in Ref. 5 and Iorio in
Ref. 6 derived an extra-secular precession of the pericentre
$\omega$ of the orbit of a test particle
\begin{equation}\dot\omega\approx\mp\rp{3c}{8r_0}\left(1-\rp{13}{32}e^2\right)\end{equation}
of $5\times 10^{-4}$ arcseconds per century ($''$ cy$^{-1}$),
while Iorio in Ref. 6 showed that also the mean anomaly
$\mathcal{M}$ is affected by DGP gravity at a larger extent
\begin{equation}\dot{\mathcal{M}}\approx\pm\rp{11c}{8r_0}\left(1-\rp{39}{352}e^2\right);\end{equation}
the longitude of the ascending node $\Omega$
is left unchanged. As a result, the mean longitude $\lambda =
\omega +\Omega
 +\mathcal{M}$, which is
a widely used orbital parameter for nearly equatorial and circular
orbits as those of the Solar System planets, undergoes a secular
precession of the order of $10^{-3}$ $''$ cy$^{-1}$. Such
precessions are independent of the semi-major axis $a$ of the
planetary orbits and depends only on their eccentricities $e$ via
second-order terms. The effects of DGP gravity on the orbital
period of a test particle were worked out by Iorio in Ref. 7;  the
DGP precession of a spin can be found in Ref. 8, but it is too
small to be detectable in any foreseeable future.

Recent improvements in the accuracy of the data reduction process
for the inner planets of the Solar System\ct{Pit05a, Pit05b},
which can be tracked via radar-ranging, have made the possibility
of testing DGP very thrilling \ct{Ior05a, Ior05b, Ior05c, Ior06a}.
In particular, Iorio in Ref. 12  showed that the recently observed
secular increase of the Astronomical Unit\ct{Kra04, Sta05} can be
explained by the self-accelerated branch of DGP and that the
predicted values of the Lue-Starkman perihelion precessions for
the self-accelerated branch are compatible with the recently
determined extra-perihelion advances\ct{Pit05b}, especially for
Mars, although the errors are still large. Rather surprisingly, it
was recently showed in Ref. 15 that the Kuiper belt objects, if
not properly modelled in the dynamical force models of the
data-reduction softwares used to process planetary data, might
affect  the dynamics of the Earth and Mars at a non negligible
level with respect to the DGP features of motion. The possibility
of using the outer planets of the Solar System, suggested by Lue
in Ref. 2 and, in principle, very appealing because all the
compering Newtonian and Einsteinian orbital effects are smaller
than the DGP precessions, is still very far from being
viable\ct{IoG06}. Finally, we mention that it was argued\ct{Ciu04}
that the launch of a LAGEOS-like Earth artificial satellite would
allow to measure the DGP perigee precession, but such a proposal
was proven to be highly unfeasible in Ref. 11.

\section*{Acknowledgements}
I am grateful to R. Ruffini and H. Kleinert for the grant received
to attend the Eleventh Marcel Grossmann Meeting on General
Relativity, 23-29 July, Freie Universit$\ddot{\rm a}$t Berlin,
2006.


\end{document}